\documentclass[aps,prl,showpacs,reprint]{revtex4-1}

\usepackage{graphicx}
\usepackage{amsmath, amsthm, amssymb}

\begin{document}

\title{Molecular Frisbee: Motion of Spinning Molecules in Inhomogeneous Fields}

\author{Johannes Flo\ss}
\author{Erez Gershnabel}
\author{Ilya Sh. Averbukh}
\affiliation{Department of Chemical Physics, The Weizmann Institute of Science, Rehovot 76100, ISRAEL}

\begin{abstract}
Several laser techniques have been suggested and demonstrated recently for preparing polarizable molecules in rapidly spinning states with a disc-like angular distribution~\cite{karczmarek98,*villeneuve00,*spanner01,*spanner01a,fleischer09,york09,kitano09,lapert09}.
We consider motion of these spinning discs in inhomogeneous fields, and show that  the molecular trajectories may be  precisely controlled by the  tilt of the plane of the laser-induced rotation.
The feasibility of the scheme is illustrated by optical deflection of linear molecules twirled by two delayed cross-polarized laser pulses.  These results  open new ways for many applications involving molecular focusing, guiding and trapping, and may be suitable for separating  molecular mixtures by optical and static fields.

\end{abstract}
\maketitle

It is a common fact that in a thermal ensemble of rotating molecules the orientation of individual molecules, as well as the orientation of their vector of angular momentum, is completely random.
Recent years have seen a tremendous progress in developing methods for the creation and fine control of non-thermal molecular angular distributions by the help of short laser pulses~\cite{stapelfeldt03}.
Several studies addressed the challenge of the optical preparation of molecules in a highly-spinning  state~\cite{karczmarek98,*villeneuve00,*spanner01,fleischer09,york09,kitano09,lapert09} with a well defined direction of the vector of  angular momentum.
At least two routes leading to this goal have been demonstrated experimentally, including the ``optical molecular centrifuge''~\cite{karczmarek98,*villeneuve00,*spanner01}, where the molecules are angularly accelerated by a laser field with a spinning polarization vector, and the ``molecular propeller'' technique~\cite{fleischer09,york09,kitano09}, which uses two time-delayed cross-polarized short laser pulses.

\begin{figure}
\centering
\includegraphics[width=0.3\linewidth]{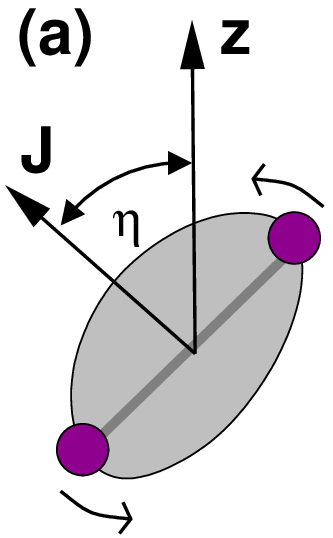}
\hspace{0.05\linewidth}
\includegraphics[width=0.6\linewidth]{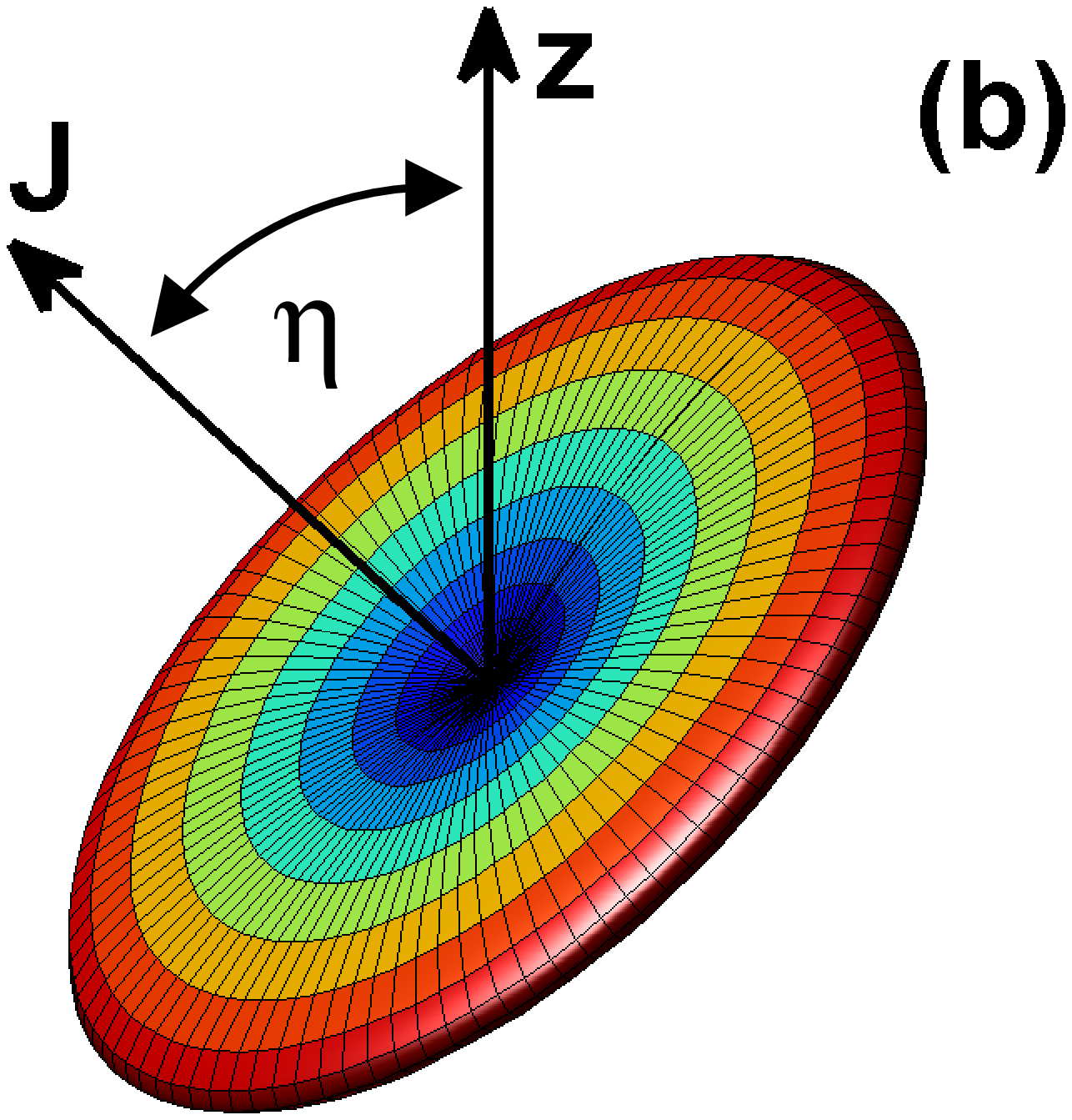}
\caption{\label{frisbee} (a) A classical molecule rotates in the plane perpendicular to its angular momentum $\mathbf{J}$.
(b) Disc-like angular distribution for the tilted spinning state $|J,J\rangle_{\eta}$ with $J=25$. The state is almost confined to a plane, as in the classical case (a).}
\end{figure}

The molecular axis of the fast-spinning molecules is confined to the rotation plane perpendicular to the vector of the angular momentum, and the molecular angular distribution has a specific disc-like shape (see Fig.~\ref{frisbee}).
In this Letter we consider deflection of the spinning molecules by inhomogeneous fields, and show that the trajectories of these ``flying discs'' may be controlled and fine-tuned by inclination of the plane of laser-induced rotation with respect to the external fields.
A similar technique is used by Frisbee players finessing  the tilt of the spinning disc for directing it into a pair of waiting hands.

In the present Letter we will concentrate on the molecular deflection by optical fields, which is a hot experimental subject~\cite{stapelfeldt97,*sakai98,bum00,*hoi01,*bum03,purcell09}, although the similar arguments are applicable to the scattering in static fields as well.
When a molecule enters a nonuniform laser field, the latter induces molecular polarization, interacts with it, and deflects the molecule along the intensity gradient.
As most molecules have an anisotropic polarizability, the deflecting force depends on the molecular orientation with respect to the deflecting field, which couples the rotational and translational motion~\cite{friedrich95,*friedrich99,purcell09,seideman97,*seideman97a,*seideman99}.
It was recently shown~\cite{gershnabel10,*gershnabel10a} that laser-induced molecular prealignment by additional short laser pulses may change dramatically the deflection process, and, in particular, reduce substantially the dispersion of the scattering angles.
Here we show that if the molecules are prepared in the fast-spinning state, not only they are scattered into a narrow angular interval, but also the position of this scattering peak is controllable by ultra-fast lasers.

For certainty, we follow a deflection scheme resembling the experiment by Stapelfeldt \textit{et al.}~\cite{stapelfeldt97} who used a strong IR laser to deflect a $CS_2$ molecular beam, and then addressed a portion of the deflected molecules (at a preselected place and time) by an additional short and narrow ionizing pulse.
Consider deflection (in the $z$ direction) of a linear molecule moving in the $x$ direction with velocity $v_x$ and interacting with a focused nonresonant laser beam that propagates along the $y$ axis.
The spatial profile of the laser electric field in the $xz$ plane is $E=E_0\exp\left[-(x^2+z^2)/\omega_0^2\right]\exp\left[-2\ln2 t^2/\tau^2\right]$.
The interaction potential of a linear molecule in the laser field is given by
\begin{equation}
U(t)=-\frac{1}{4}E^2\left(\Delta\alpha\cos^2\theta+\alpha_{\perp}\right)\,,
\label{eq6}
\end{equation}
where $E$ is defined above, and $\Delta\alpha\equiv\alpha_{\parallel}-\alpha_{\perp}$, where $\alpha_{\parallel}$ and $\alpha_{\perp}$ are the molecular polarizabilities along and perpendicular to the molecular axis, respectively.
Here $\theta$ is the angle between the electric field polarization direction (along the laboratory $z$ axis) and the molecular axis.
A molecule initially moving along the $x$ direction will acquire a velocity component $v_z$ along the $z$ direction.
In the weak field regime, when the corresponding deflection angle $\gamma\approx v_{z}/v_{x}$ is small, we substitute $x=v_x t$, and consider $z$ as a fixed impact parameter.
The deflection velocity is given by
\begin{equation}
v_{z}=-\frac{1}{M}\int_{-\infty}^{+\infty}\left(\vec{\nabla}U(t)\right)_{z}dt\ ,
\label{eq7}
\end{equation}
where $M$ is the mass of the molecule. The time dependence of the potential in Eq.~\ref{eq7} comes from three sources:
pulse envelope, projectile motion of the molecule through the laser focal area, and time variation of the angle $\theta$ due to molecular rotation.
For simplicity, we assume also that the  deflecting field does not affect significantly the rotational motion.
Such an approximation is justified, say for $CS_2$ molecules with the rotational temperature $T=5~\text{K}$, which are subject to a deflecting field of $3\times 10^9~\text{W}/\text{cm}^2$.
Since the rotational time scale is the shortest one in the problem, we average the force over the fast rotation.
Considering molecular rotation classically (see Fig.~\ref{frisbee}a),  we arrive at the following expression for the deflection angle, $\gamma= v_{z}/v_{x}$~\cite{gershnabel10a}:
\begin{equation}
\gamma=\gamma_0/\overline{\alpha}\left(\Delta\alpha\mathcal{A}+\alpha_{\perp}\right) \,.
\label{eq1}
\end{equation}
Here, $\overline{\alpha}=1/3\alpha_{\parallel}+2/3\alpha_{\perp}$ is the orientation-averaged molecular polarizability, $\gamma_0$ is the average deflection angle for an isotropic molecular ensemble, which is determined by the deflection scheme~\cite{stapelfeldt97,*sakai98}, and $\mathcal{A}=\overline{\cos^2\theta}$ is the time-averaged value of $\cos^2\theta$.
This quantity depends on the relative orientation of the vector of angular momentum and the polarization of the deflecting field.
It is different for different molecules of the incident ensemble, which leads to the randomization of the deflection process.
Qualitatively, the properties of ${\cal A}$ may be understood from the following classical arguments.
Consider a linear molecule that rotates freely in a plane that is perpendicular to the vector $\mathbf{J}$ of the angular momentum (see Fig.~\ref{frisbee}a).
The projection of the molecular axis on the vertical $z$ direction is given by $\cos\theta (t)=\cos(\omega t)\sin \eta$, where $\eta$ is the angle between $\mathbf{J}$ and the deflecting field, and $\omega$ is the angular frequency of molecular rotation. Averaging over time, one arrives at
\begin{equation}
\mathcal{A}=\overline{\cos^2\theta}=\frac{1}{2}\sin^2\eta.
\label{eq8}
\end{equation}
In a thermal ensemble, vector $\mathbf{J}$ is randomly oriented in space, with isotropic angular distribution.
Therefore, the corresponding distribution of ${\cal A}$ is~\cite{gershnabel10}
\begin{equation}
f({\cal A})=\frac{1}{\sqrt{1-2{\cal A}}},\label{rainbow}
\end{equation}
which has a rainbow singularity at ${\cal A}=1/2$ and a flat step near the minimal ${\cal A}=0$.

However, for fast-spinning molecules the distribution of $\mathbf{J}$ is completely different, as $\mathbf{J}$ is mostly perpendicular to the well-defined molecular plane of rotation. Since $\eta$ takes a well-defined value,  we expect the distribution of ${\cal A}$ to be very narrow and, more importantly, to be centered around the value determined by $\eta$.

For a more quantitative quantum treatment, we consider a linear molecule in a tilted spinning state $|J,m=J\rangle_{\eta}$, where $J$ is the angular momentum quantum number and $m$ is the quantum number associated with the projection of angular momentum on the tilted $\mathbf{J}$ direction (see Fig.~\ref{frisbee}b).
The angular distribution for a molecule in the  $|J=25,m=25\rangle_{\eta}$ state is plotted in Fig.~\ref{frisbee}b, where we observe that the molecule is indeed mostly confined to the plane perpendicular to the direction of $\mathbf{J}$.
The generation of such a state is feasible, e.g. by the optical centrifuge approach~\cite{karczmarek98}, hexapole selection~\cite{baugh94}, or selection by deflection~\cite{Holmegaard09,*Filsinger09}.

To obtain the distribution of $\mathcal{A}$ for a molecule being in such a state, we expand  $|J,J\rangle_{\eta}$ in the basis of $|J,m\rangle$ states with the quantization axis parallel to the deflecting field.
This is essentially obtained by rotating the state by the angle $\eta$ (see Fig.~\ref{frisbee}b):
\begin{equation}
|\Psi\rangle=\hat R (\eta) |J,J\rangle_{\eta} = \sum_{m=-J}^{+J}c_{J,m}|J,m\rangle\ ,
\label{eq3}
\end{equation}
where $\hat R (\eta)$ is the rotation matrix~\cite{CT}.
The expansion coefficients $c_{J,m}$ are given by~\cite{steinborn73}
\begin{equation}
c_{J,m}=\sqrt{\left(\begin{array}{c}2J\\J-m\end{array}\right)}\left(\cos\frac{\eta}{2}\right)^{J+m}\left(-\sin\frac{\eta}{2}\right)^{J-m} \,.
\label{eq4}
\end{equation}

For a weak enough deflecting field, the scattering angle for a molecule in the $|J,m\rangle$ rotational state is given by Eq.~\ref{eq1}, in which ${\cal A}$ is replaced by
\begin{eqnarray}
\mathcal{A}_{J,m}&=& \langle J,m|\cos^2\theta|J,m\rangle \nonumber\\  &=&\frac{1}{3}+\frac{2}{3}\frac{J(J+1)-3m^2}{(2J-1)(2J+3)}\,.
\label{eq2}
\end{eqnarray}

\begin{figure}
\centering
%\includegraphics[width=\linewidth]{deflectionjjstateDiscrete}
%\hspace{0.05\linewidth}
%\includegraphics[width=\linewidth]{deflectionjjstate400.eps}
\includegraphics[width=\linewidth]{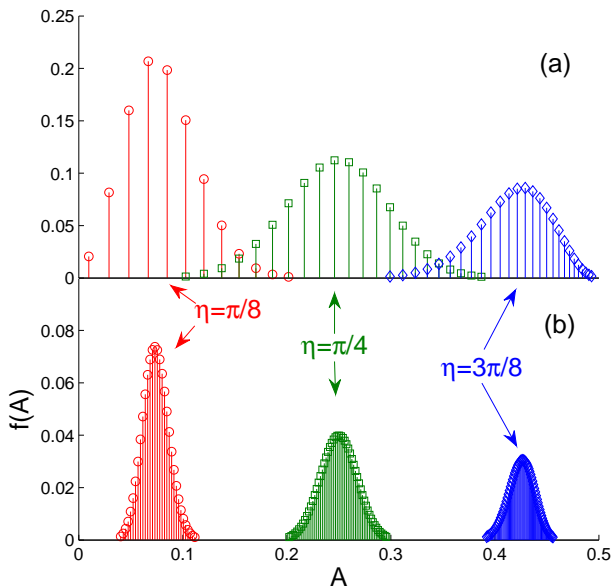}
\caption{\label{deflectionjjstate} (a) Distribution of $\mathcal{A}$ for a $|J,J\rangle_{\eta}$ state with $J=50$ and different angles $\eta$ between $\mathbf{J}$ and the deflecting field.
(b)~Same as (a) but for $J=400$.
The distribution of $\mathcal{A}$ is rather narrow, and by adjusting the angle $\eta$, its maximum can be tuned between $0$ and $0.5$. The maximum of each curve is approximately at $1/2\sin^2\eta$ (Eqs.~\ref{eq8} and~\ref{eq21}).}
\end{figure}

Fig.~\ref{deflectionjjstate} shows the distribution of ${\cal A}_{J,m}$ values weighted by the population $|c_{J,m}|^2$ of each state (for several values of $\eta$). Since the relation between $m$ and ${\cal A}_{J,m}$ is known (Eq.~\ref{eq2}, where $J$ is a parameter), we obtain
\begin{equation}
f(\mathcal{A})=\sum_{i=1,2} \left|\frac{d m}{d \mathcal{A}}\right| f(m)^{(i)} \,,
\label{eq20}
\end{equation}
where the summation is over two branches of the function $m(\mathcal{A})$.
Furthermore, we note that $f(m)=|c_{J,m}|^2$ is a binomial distribution, which becomes a Gaussian distribution in the limit of large $J$.
Therefore, for $J\gg 1$ and $0< \eta< \pi$ the distribution of ${\cal A}_{J,m}$ for a molecule prepared in the $|J,J\rangle_{\eta}$ state is
\begin{eqnarray}
f(\mathcal{A})&=&\frac{1}{\sqrt{2\pi\sigma^2}}\frac{1}{\sqrt{1-2A}}\nonumber\\
&\times&\sum_{i=1,2}\exp\left[-\frac{\left(\cos\eta+(-1)^i\sqrt{1-2A}\right)^2}{2\sigma^2}\right],
\label{eq21}
\end{eqnarray}
with $\sigma=|\sin\eta|/\sqrt{2J}$.
The maximum of the distribution is located near $\mathcal{A}_{max}=1/2\sin^2\eta$, which is exactly the classical value for $\mathcal{A}$ (see Eq.~\ref{eq8}).
Hence, preparing a molecule in a $|J,J\rangle_{\eta}$ state (with the appropriate $\mathbf{J}$ direction) allows for controlling the deflection angle.
The corresponding distribution of the scattering angle is much more localized compared to the thermal distribution (Eq.~\ref{rainbow}) which is spread out over the entire ${\cal A}_{J,m}$ range.

To analyze the experimental feasibility of the suggested control scheme, we consider optical deflection of unidirectionally rotating molecules prepared by the double-pulse ``molecular propeller'' technique~\cite{fleischer09,york09,kitano09}.
Motivated by the so-far success in predicting the main results classically, we provide here only classical calculations.
Full quantum mechanical treatment which shows a strong agreement with the classical results will be published elsewhere~\cite{floss10a}.
Let us recall briefly the corresponding scenario for inducing molecular spinning (a detailed discussion of this scenario can be found in Refs.~\cite{fleischer09,khodorkovsky10}).
The first short linearly polarized laser pulse (with the polarization vector $\mathbf{p}_1$) brings molecules to a concerted rotation that results in molecular alignment along the $\mathbf{p}_1$-direction sometime after the end of the pulse (i.e. at field-free conditions).
At the moment of the maximal alignment, when the molecular angular distribution is mostly confined in a narrow cone around $\mathbf{p}_1$, the second, delayed laser  pulse is applied, with the polarization vector $\mathbf{p}_2$ at $45$ degrees with respect to the first one. As a result, the molecules gain an angular momentum in the $\mathbf{p}_1\times\mathbf{p}_2$ direction, and start spinning in the plane spanned by the vectors $\mathbf{p}_1$ and $\mathbf{p}_2$.

We performed Monte Carlo simulations for a thermal molecular ensemble excited by the above double-pulse sequence for various orientations of the plane of induced spinning with respect to the deflecting field.
Since the needed laser pulses are much shorter than the rotational time-scale, we treated them as $\delta$-kicks.
The change of the angular momentum of individual molecules due to a pulse kick  is given by~\cite{khodorkovsky10}
\begin{equation}
\mathbf{\Delta J} = 2\hbar P \left(\mathbf{p} \cdot \mathbf{r} \right) \mathbf{r} \times \left[\mathbf{p}-\left(\mathbf{p}\cdot\mathbf{r}\right)\mathbf{r}\right]\,,
\label{eq5}
\end{equation}
where $\mathbf{p}$ is the unit polarization vector of the laser pulse and $\mathbf{r}$ is unit vector along the molecular axis at the moment of the pulse.
The parameter $P=\Delta\alpha/(4\hbar)\int E^2(t) dt$ is an effective kick-strength, characterizing the laser-induced angular momentum influx (in the units of $\hbar$).
After calculating the final value of $\mathbf{J}$ for every molecule after the two pulses, we obtained $\mathcal{A}$ from Eq.~\ref{eq8} by using $\sin^2\eta=1-(\mathbf{J}\cdot\mathbf{d})^2/J^2$, where $\mathbf{d}$ is the unit vector along the deflecting field.

\begin{figure}
\includegraphics[width=\linewidth]{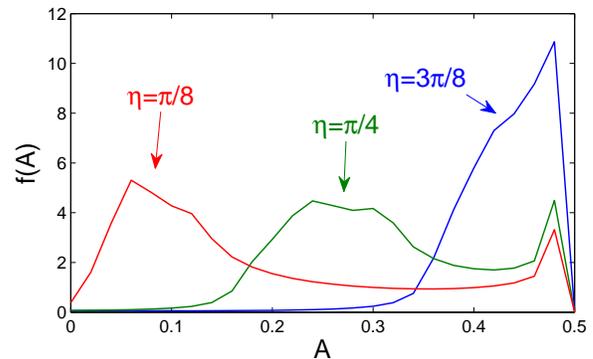}
\caption{\label{deflectionclassical}Distribution of $\mathcal{A}=\overline{\cos^2\theta}$ for a classical thermal ensemble of molecules ($J_T=5$) on a coarse-grained grid.
The molecules are excited by two delayed laser pulses cross-polarized at 45 degrees with respect to each other.
The kick-strengths of the pulses are $P_1=10$ and $P_2=50$.
}
\end{figure}

The resulting distribution of $\mathcal{A}$ is shown in Fig.~\ref{deflectionclassical} for various values of the angle $\eta$.
The parameters of the presented simulation are $P_1=10$, $P_2=50$ and $J_T=5$.
Here $J_T=\sqrt{k_B T /(2B)}$ is a typical ``thermal'' value of $J$ (for $J_T\geq 1$), where $T$ is the temperature, $k_B$ is Boltzmann's constant, and $B$ is the rotational constant of the molecules.
In the case of $CS_2$ molecules, the values of $P=10,50$ correspond to the excitation by $0.5~\text{ps}$ (FWHM) laser pulses with the maximal intensity of $1.1 \times 10^{12}~\text{W}/\text{cm}^2$ and $5.6 \times 10^{12}~\text{W}/\text{cm}^2$, respectively, and $J_T=5$ corresponds to $T=3.9~\text{K}$.
Figure~\ref{deflectionclassical} clearly shows that the thermal distribution of the deflection angles (see Eq.~\ref{rainbow}) is drastically modified by the laser-induced spinning.
Distinguishable condensed deflection peaks with a tunable position are indeed formed, as predicted above.
The distribution elevation near $\mathcal{A}=0.5$ is due to an integrable singularity (similar to the one described by Eq.~\ref{rainbow}), which is caused by a residual fraction of molecules with $\eta$ close to $\pi/2$ because of the non-perfect confinement of the molecules to the rotation plane.
To show the distribution of $\mathcal{A}$ unobstructed by this singularity, we used a coarse-grained bin averaging in Fig.~\ref{deflectionclassical}.

Our results demonstrate that the predicted effect is clearly observable using a rather basic double-pulse scheme for spinning molecules~\cite{fleischer09,york09,kitano09}, which was demonstrated in a recent experiment~\cite{kitano09}.
An even better outcome can be achieved by the optical centrifuge approach~\cite{karczmarek98,*villeneuve00,*spanner01} which has the potential for generating clean $|J,J\rangle$ states.
A recent proposal~\cite{lapert09} for generating the coherent superposition $1/\sqrt{2}(|J,J\rangle + |J,-J\rangle)$ by the help of properly shaped perpendicularly polarized laser fields is instrumental for our purposes as well.
We hope that the results presented in this Letter will encourage further development of those schemes.

Summarizing, we have shown that a high degree of control over the molecular motion in inhomogeneous fields is available by preparing the molecules in highly spinning states with a tunable tilt of the plane of rotation.
Controlling deflection direction, and narrowing the distribution of the deflection angle is important for nanofabrication schemes based on the molecular optics approach~\cite{seideman97}.
Moreover, molecular deflection by nonresonant optical dipole force is considered as a promising route to the separation of molecular mixtures (for a recent review, see~\cite{Zhao06}).
Isotope-selective spinning was already demonstrated in the optical centrifuge~\cite{karczmarek98,*villeneuve00,*spanner01}, and it may be integrated with the present technique for a non-destructive separation process. Furthermore, other existing laser methods for selective alignment of molecular isotopes~\cite{fleischer06}, or nuclear spin isomers~\cite{Renard04,fleischer07,*Gershnabel08}, can be extended to selective spinning, and used for separation purposes as well.
The Frisbee-like game with spinning molecules may open new ways for many applications involving molecular focusing, guiding, and trapping by optical and static fields.

We appreciate many fruitful discussions with Yehiam Prior, Sharly Fleischer and Yuri Khodorkovsky. Financial support of this research by the Israel Science Foundation is gratefully acknowledged.  The work of JF is supported by the Minerva Foundation. IA is an incumbent of the Patricia Elman Bildner Professorial Chair. This research is made possible in part by the historic generosity of the Harold Perlman Family.

%\bibliography{bibliographyFrisbee}
%merlin.mbs 2010-03-15 4.21a (PWD, AO, DPC)
%Control: key (0)
%Control: author (8) initials jnrlst
%Control: editor formatted (1) identically to author
%Control: production of article title (-1) disabled
%Control: page (0) single
%Control: year (1) truncated
%Control: production of eprint (0) enabled
%

\end{document}